# Targeting DNA methylation: new paradigms and the advent of gene-selective tools

Julie Gilbert[1] and Francesco Calzaferri[2]*

[1] University of Oxford, Department of Chemistry, 12 Mansfield Road, Oxford, OX1 3TA, United Kingdom

[2] Institut des Biomolécules Max Mousseron (IBMM, UMR 5247 CNRS-UM-ENSCM), 1919 Route de Mende, 34293 Montpellier, France

*Corresponding author; email: francesco.calzaferri@cnrs.fr

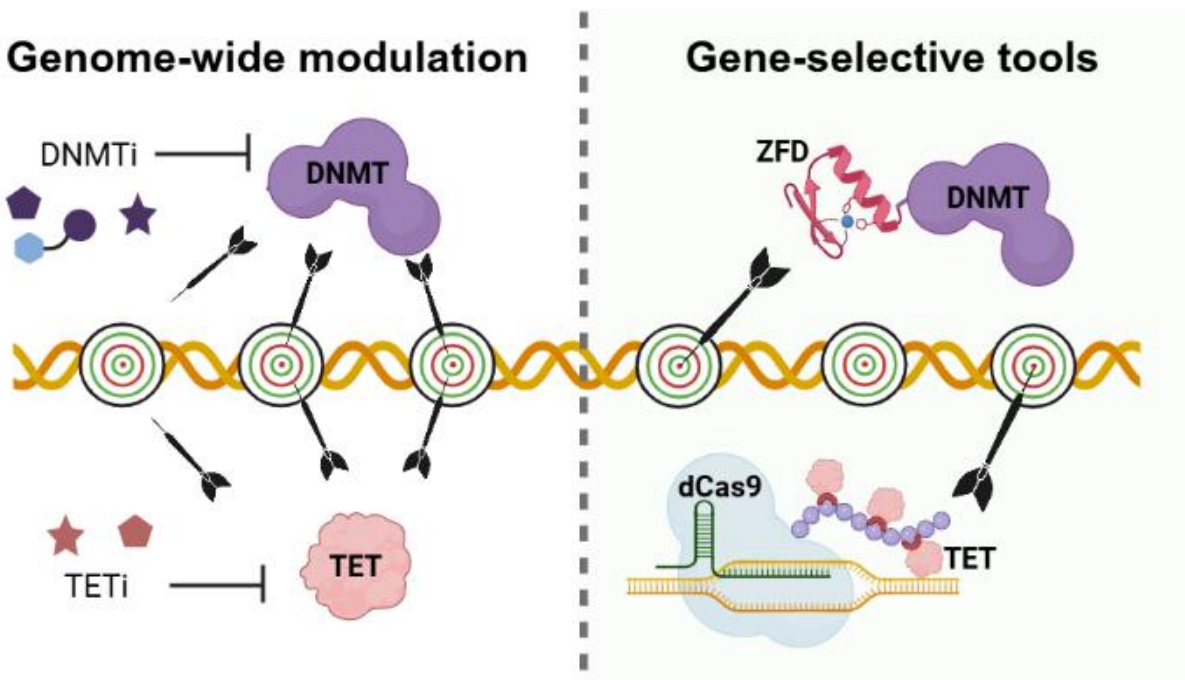

## Abstract

DNA methylation can function as a toxic alkylation reaction exploited by chemotherapeutic agents to induce cancer cell death. However, finely tuned DNA methylation plays a fundamental role in cellular physiology, particularly in the epigenetic regulation of gene expression. Once thought to act solely as a repressor of gene transcription, its functional role has since been elucidated as genomic locus-specific and deeply connected with other epigenetic factors. Following the clinical approval of DNA methyltransferase inhibitors, such as Azacitidine and Decitabine, for the treatment of haematological malignancies, considerable efforts have been devoted to developing pharmacological tools that modulate epigenetic DNA methylation. However, the lack of gene selectivity in these agents limits their therapeutic efficacy and increases off-target toxicity. Moreover, the non-gene-selective nature of current DNA methylation-targeting molecules fails to meet the standards required to discern the nuanced roles of DNA methylation across diverse pathophysiological contexts and genomic loci, particularly in an era where next-generation sequencing and omics technologies enable high-resolution epigenetic analyses. In this review, we examine the mechanisms and roles of DNA methylation in epigenetic regulation, evaluate the current landscape of DNA methylation modulators, from traditional DNMT inhibitors to cutting-edge CRISPR-dCas9 fusion systems and protein-protein interaction disruptors, and discuss their clinical relevance. Finally, we emphasise the need for precise, locus-specific tools to advance both cancer research and therapeutic strategies.

## 1. Introduction

The term "DNA methylation" chemically refers to the addition of a methyl group to DNA nucleobases, a process known as an alkylation reaction. In toxicology, chemically induced DNA methylation is primarily associated with deleterious effects, including mutagenesis and carcinogenesis, due to DNA damage. More broadly, alkylating agents exploit this toxic mechanism to trigger cancer cell death during chemotherapy. This represents the primary mechanism of action for classic chemotherapeutic agents, such as nitrogen mustards (e.g., Melphalan, Chlorambucil)[1] and nitrosoureas (e.g., Lomustine, Carmustine).[2] In particular, temozolomide (TMZ), another anticancer agent used in the treatment of glioblastoma, introduces methyl groups onto DNA in a non-specific manner, with a preference for guanine at positions N7 and O6, and adenine at position N3.[3,4] When the DNA repair machinery fails to intervene effectively, these modifications can induce DNA mismatches, single- and double-strand breaks, ultimately leading to cell cycle arrest and cell death (**Figure 1**).[5]

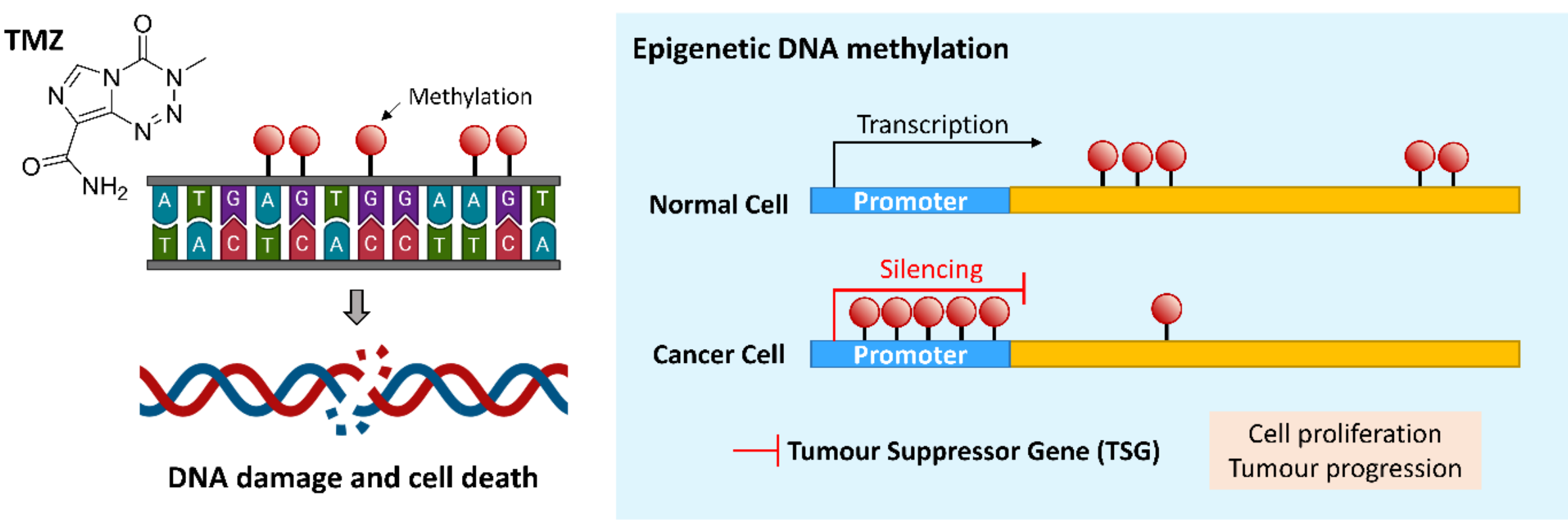

**Figure 1**. Representation of the mechanism of action of temozolomide (TMZ) as representative alkylating agent used in cancer therapy (left); traditional paradigm of the function of epigenetic DNA methylation in normal and cancer cells (right).

However, DNA methylation is also an essential physiological mechanism used by cells to regulate gene expression in a reversible and heritable manner. This role is shared with other mechanisms occurring on the chromatin, including histone modifications such as methylation and acetylation, as well as the action of ATP-dependent chromatin remodelling factors like the SNF2 and SWI proteins, and non-coding RNAs, which are gathered together under the name of "epigenetic mechanisms".[6] When we talk about DNA methylation as an epigenetic modification, we refer to the methylation at position 5 of cytosines followed by guanosines (CpG) in the DNA primary sequence of mammals, a process catalysed by DNA methyltransferases (DNMTs). Methylation of cytosines followed by other nucleobases (mCH) has also been detected in vertebrates, albeit in low abundance (1-3%), predominantly in oocytes, neural tissues, and embryonic stem cells.[7] While well-established in bacteria,[8] methylation of other nucleobases, such as adenine at position N6, has also been detected in mammals, although its functional role remains less understood.[9,10] For this reason, throughout this review, the term "DNA methylation" will refer to methylation at CpG sites unless otherwise specified.

CpG sites can accumulate in specific genomic regions. According the original definition, these regions are called CpG islands (CGIs) when the CpG content exceeds 50% in a sequence of at least 200-1000 base pairs (bp), and the ratio of observed to expected CpGs (defined as the number of CpG sites over the number of C and G nucleotides taken separately) is greater than 0.6.[11] This definition remains valid in the field and is used to bioinformatically identify CGI in the genome. Approximately 70% of annotated gene proximal promoters, which are regulatory regions for gene transcription initiation, overlap with a CGI,[12] underscoring the role of DNA methylation in gene expression.

Over the past few decades, epigenetic DNA methylation has been extensively studied, revealing its significant contributions to early development and the progression of various pathologies, including viral infections, neurodegeneration, and cancer.[13–15] Additionally, differential methylation patterns on specific genes in cancer versus normal cells are now being considered as potential biomarkers for early cancer detection.[16] The therapeutic efficacy of targeting aberrant DNA methylation has been validated by the clinical approval of nucleoside DNMT inhibitors (DNMTi), such as Azacitidine and Decitabine, for the treatment of myelodysplastic syndromes, chronic myelomonocytic leukemia, and acute myeloid leukemia.[17,18] For this reason, research has focused more on the field of cancer treatment than other disorders, but new avenues must still be explored to understand further the implication and therapeutic potential of DNA methylation.

Traditionally, DNA methylation of gene promoters leads to gene silencing and the clinical benefits associated with DNMTi in cancer were linked to the re-expression of tumour suppressor genes (TSGs) that were silenced in malignant cells (**Figure 1**).[19] Recently, the advent of next-generation sequencing and omics techniques, such as methylomics and transcriptomics, has deepened our understanding of DNA methylation not only on gene promoters and transcription sites, but also on regulatory sites within gene bodies, transposomes, and repetitive regions.[13] This has revealed the DNA methylation landscape to be more complex and fascinating than previously anticipated, with many more roles and nuances yet to be uncovered. For chemical biologists studying this epigenetic mechanism, it has become clear that developing tools to modulate DNA methylation at specific genomic loci, rather than globally, as traditional DNMTi do, is the current goal. Although far from clinical relevance, scientists have begun developing such tools over the past decade.[20,21] In this review, we aim to untangle the necessity of developing efficient tools to control DNA methylation in a locus-specific manner, addressing both fundamental research and potential therapeutic applications. We explore the mechanisms and roles of DNA methylation in epigenetic regulation, examine the clinical aspects of current DNA methylation modulators, and highlight the emerging potential of gene-selective probes to overcome the limitations of non-specific approaches.

## 2. Endogenous DNA methylation mechanism in cells

In mammals, DNA methylation is catalysed by DNMT1, DNMT3A, and DNMT3B, which convert cytosines into 5-methylcytosine (5mC). All the three enzymes share a conserved C-terminal methyltransferase (MTase) catalytic domain that facilitates the formation of a ternary complex with DNA and S-adenosyl-L-methionine (SAM), the universal methyl donor (**Figure 2**). The targeted cytosine flips out of the DNA double strand and is stabilized by glutamic acid within the catalytic pocket. The catalytic cysteine of the MTase (C1229 in DNMT1, C710 in DNMT3A, and C657 in DNMT3B) then binds to the C6 position via Michael addition, forming a transient covalent intermediate that activates the C5 position, and facilitates the transfer of the methyl group from SAM. S-adenosyl-L-homocysteine (SAH) is subsequently released, and β-elimination of the cysteine allows the reinsertion of the methylated cytosine into the DNA helix.[22,23]

*A contrario*, DNA methylation can be reversed through passive and active demethylation. Passive demethylation, also known as "dilution", occurs when a the 5mC mark is not maintained during DNA replication.[13] Active demethylation is driven by the ten-eleven translocation (TET) family of dioxygenases, which includes TET1, TET2, and TET3. These enzymes catalyse the consecutive oxidation of 5mC to 5-hydroxymethylcytosine (5hmC), 5-formylcytosine (5fC), and 5-carboxylcytosine (5caC) using Fe(II), 2-oxoglutarate, and dioxygen as cofactors. Oxidised cytosines alter DNMT recognition during DNA replication and the binding of other epigenetic partner proteins, leading to a loss of methylation maintenance across cell division or through the engagement of DNA repair mechanisms. Indeed, 5fC and 5caC are replaced by unmethylated cytosine through base excision repair (BER).[24–26]

During mammalian embryogenesis, the methylome is completely erased after fertilization and during gametogenesis, and subsequently reprogrammed through *de novo* methylation by DNMT3A and DNMT3B.[27,28] The newly established methylation patterns are then maintained during replication by DNMT1, ensuring their inheritance during cell division (**Figure 3A**).[29] These observations date back to more than 30 years ago and defined the first model of *de novo* and maintenance DNA methylation. Depending on lifestyle and environmental exposures, methylation patterns evolve through the cooperation and synergy of all three enzymes.

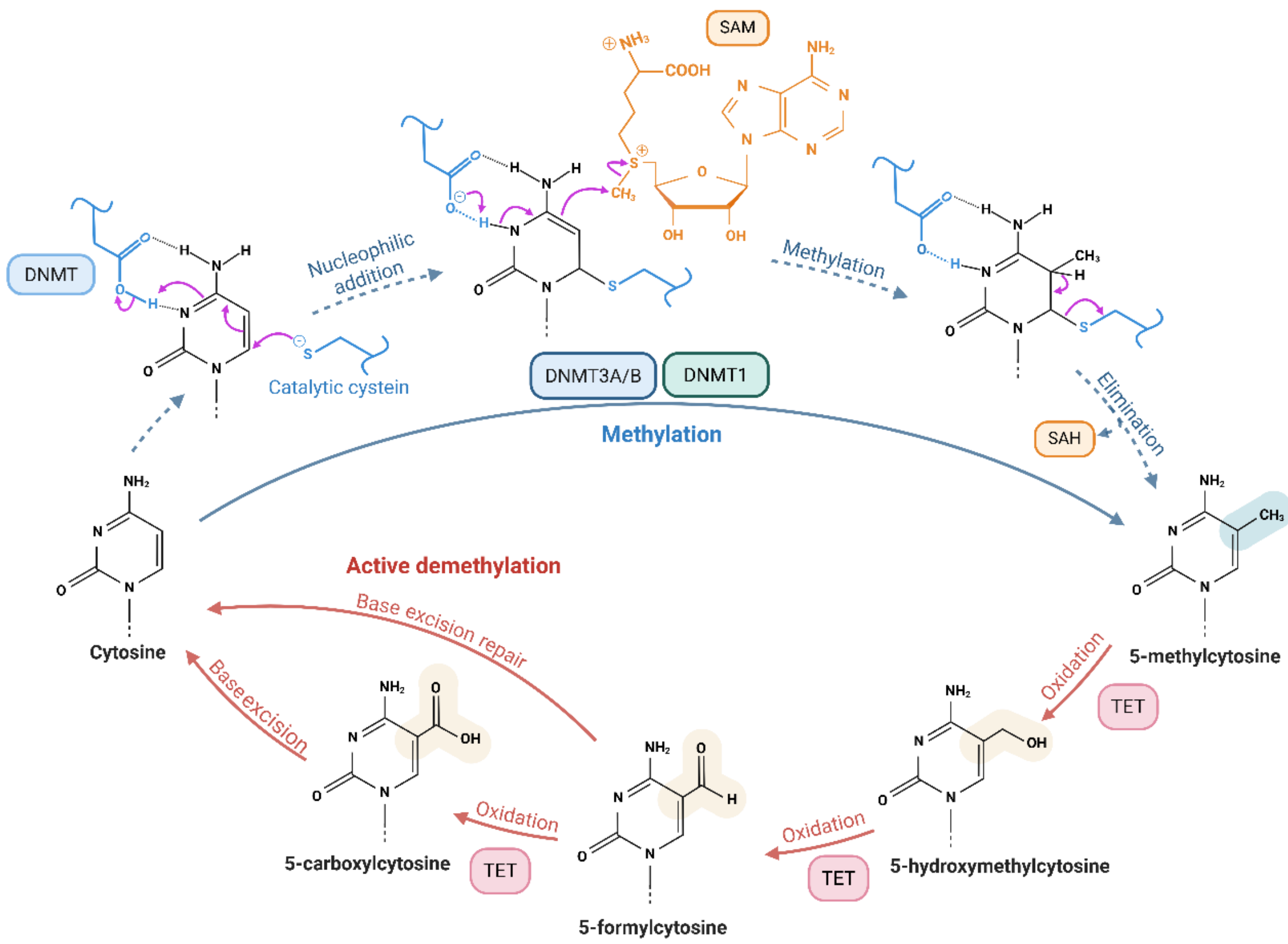

**Figure 2**. DNMT-driven DNA cytosine methylation and active demethylation in cells. The cytosine is methylated at position 5 by DNMTs in the presence of the co-factor SAM through a mechanism involving a DNMT catalytic cysteine addition, methyl transfer, and enzyme elimination. The active demethylation is mediated by the successive oxidation of the methyl by TET enzymes followed by base excision repair.

DNMT1 is the largest of these proteins, with a sequence of approximately 1600 amino acids (**Figure 3B**). In addition to the MTase domain, it comprises five domains, including the replication foci-targeting sequence (RFTS) and the bromo-adjacent homology (BAH) domains. In its free form, DNMT1 is autoinhibited by the interaction of the RFTS domain with the catalytic domain.[30] During the S phase of the cell cycle, which corresponds to DNA replication, DNMT1 levels significantly increase.[31] DNMT1 is directed to the replication fork by proliferating cell nuclear antigen (PCNA) and the ubiquitin-like protein containing PHD and RING finger domains 1 (UHRF1) through RFTS recognition.[32,33] UHRF1 guides DNMT1 to hemimethylated DNA in proximity to H3K9me2 and H3K9me3 marks,[34,35] and RFTS binds to ubiquitinated H3 tails, activating DNMT1 catalytic activity on newly synthesized DNA.[36] It has also been shown that the interaction of the BAH1 domain with the histone mark H4K20me3 is essential to displace the autoinhibitory linker, highlighting the crosstalk between epigenetic marks.[37]

DNMT3A and DNMT3B share homology with the non-enzymatic DNMT3-like protein (DNMT3L), which cooperates at the C-terminus to form the DNMT3L-DNMT3A/B-DNMT3A/B-DNMT3L heterotetrameric complex (**Figure 3A**). The homodimerization of DNMT3 is essential for recognising unmethylated DNA double strands, and interaction with DNMT3L stimulates methylation and enhances processivity.[38–40]

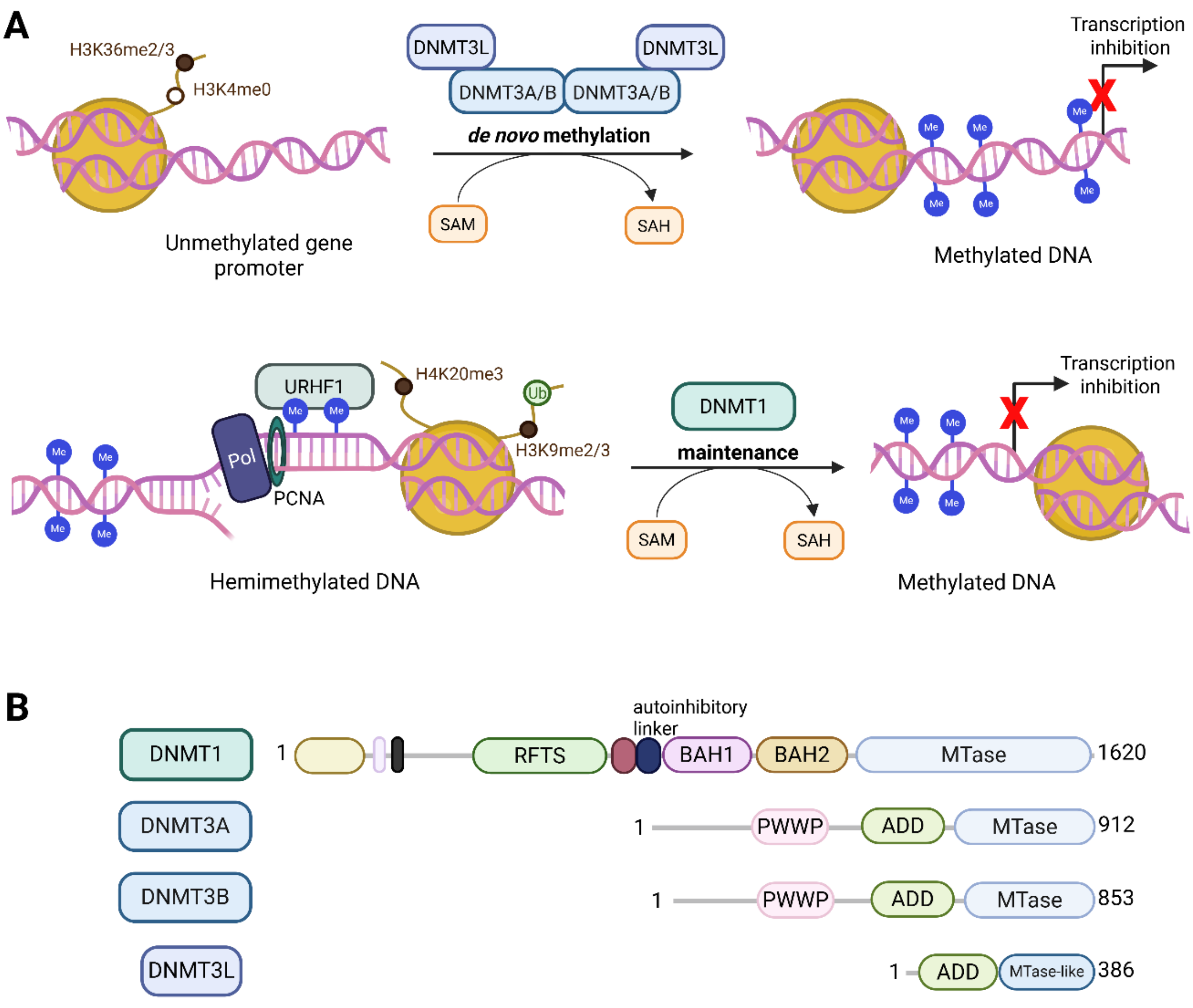

**Figure 3.** Illustration of *de novo* and maintenance DNA methylation in the regulation of gene expression. **A)** In the traditional model, *de novo* methylation on unmethylated DNA is mediated by DNMT3A/B that tetramerise with DNMT3L, wherease maintenance methylation is achieved by DNMT1 during replication at hemimethylated positions in the presence of URFH1 and PCNA. Both mechanisms involve the recognition of histone marks and the co-factor SAM, especially on gene promoters, leading to transcriptional repression. **B)** Schematic representation of the different domains of mammalian

DNMT subtypes. DNMT3L does not have the MTase catalytic domain, thus not functioning as a methyltransferase.

This was validated by crystal structures of the complexes for DNMT3A[41] and later for DNMT3B.[42] The heterotetrameric complex is thought to stabilise DNMT3A–3L, prevent higher-order oligomerization, and is considered the most relevant active form of DNMT3 organization, although tetramers can form filaments along DNA in CGI-rich regions.[43] The tetramer requires a nucleosome to methylate DNA,[44] and this binding does not require a CpG motif, indicating that it is not sequence-specific.[45] The distinct activities and roles of DNMT3A and DNMT3B remain unclear but appear correlated with their different nuclear locations[27] and distinct DNA sequence preferences, such as cytosines followed by two guanines (CpGpG sequence) for DNMT3B.[46,47] Similar to DNMT1, DNMT3A, DNMT3B, and DNMT3L interact with histone modifications, particularly with H3 tails. For example, the binding of unmethylated H3K4 (H3K4me0) to the ATRX-DNMT3-DNMT3L (ADD) domain of DNMT3 induces a conformational change that activates the enzyme's catalytic site (**Figure 3B**).[48] H3K36me2/3 also activates and stimulates DNA methylation at the amino-terminal Pro–Trp–Trp–Pro (PWWP) domain, facilitating locus recruitment.[49] However, H3K36me2 interacts exclusively with DNMT3A, while the trimethylation mark interacts with both enzymes, with a preference for DNMT3B.[50,51]

Cooperativity between *de novo* and maintenance methylation has been previously observed. For example, at transposable elements, mobile genetic regions that can change their position within the genome, both non-methylated DNA strands are not simultaneously methylated by DNMT3B alone but through cooperation between DNMT3B and DNMT1.[52,53] However, the traditional model assigning *de novo* and maintenance methylation roles to DNMT3 and DNMT1, respectively, was challenged over 20 years ago,[32] and cumulative evidence supporting this model has since been gathered.[54,55] This is not surprising given the importance of DNA methylation maintenance in cells, particularly since DNMT1 is not dispensable in somatic or cancer cells, and for embryonic stem cells growth and differentiation.[44,56,57] It is thus difficult to believe that DNMT1 alone can guarantee the proper transmission of methylation across generations.

The exchangeable roles of DNMT subtypes in maintenance and *de novo* methylation have been highlighted at specific genomic loci. For instance, *de novo* methylation by DNMT1 has been detected on repetitive elements through high-resolution sequencing[54] and even on gene promoters in cancer cell lines.[58] Additionally, maintenance methylation by DNMT3 has been postulated following observations of a 30% increase in hemimethylated repeats and a general decrease in methylation in DNMT3A and DNMT3B knockout cells.[32] This raises an intriguing question in the field: are *de novo* and maintenance methylation activities more dependent on genomic position than on the intrinsic enzymatic nature of DNMT subtypes? This hypothesis aligns with the fact that DNMTs are typically not free active enzymes in the cellular environment. DNMT1 is activated by UHRF1 at regions where H3K9 is di- or trimethylated and where H3 tails are ubiquitinated,[28] while DNMT3A and DNMT3B require polymerisation and nucleosome binding to methylate DNA.[32,59] This suggests that multiple epigenetic factors, such as histone modifications and chromatin remodelers, control DNMT recruitment to specific genomic regions, potentially having a more significant impact on their function than their ability to distinguish between hemimethylated and unmethylated sequences.[55,60] In summary, the recruitment of DNMTs by specific epigenetic partners or marks at particular genomic regions may be fundamental to elucidating the roles of DNMT subtypes.

## 3. Role of DNA methylation in gene expression

DNA methylation is generally associated with gene silencing. This association is based on evidence that i) most CGIs at gene promoters remain unmethylated in somatic cells, exposing transcription start sites for transcription factor binding, and ii) methylated promoter CGIs are typically restricted to genes with long-term repression, such as those on the inactive X chromosome, imprinted genes where only one parental allele is expressed, and germline-specific genes.[44] Additionally, the promoters of TSGs silenced in cancer, such as *VHL*, *p16$^{INK4a}$*, *p14$^{ARF}$*, *CDH1*, *CDH13*, *MLH1*, *MGMT*, *DAPK1,* and *BRCA1*, are often hypermethylated (**Figure 4**).[61,62]

However, the DNA methylation landscape is far more complex. Many cancers exhibit global hypomethylation despite hypermethylation of TSG promoters, a phenomenon termed the "DNA methylation paradox".[61] This paradox is counterintuitive given that DNMTs are frequently overexpressed in tumours, including colorectal, bladder, kidney, pancreatic cancers, and glioblastoma.[33,63] Over the past decade, it has become evident that DNA methylation exerts distinct roles depending on the targeted genomic region. For example, high intragenic methylation is associated with active gene transcription,[44,64] even when accompanied by H3K9me3 and methyl-CpG-binding protein 2 (MeCP2) binding, both of which are transcriptional repressors at gene promoters.[65] Thus, the impact of DNA methylation on gene transcription may depend more on the specific genomic regions targeted and the interplay of other epigenetic mechanisms in those regions, rather than on the intrinsic role of DNA methylation alone, as previously assumed. This is even clearer if we consider the proposed role of mCHs, usually in a CAG or CAC context. For example, neuronal mCHs are predominantly present in gene bodies and anticorrelate with gene expression.[7] Interestingly, MeCP2 is the only reader of mCH to date and it maintains its silencing role in the brain, contrary to what mentioned above for mCG.[66]

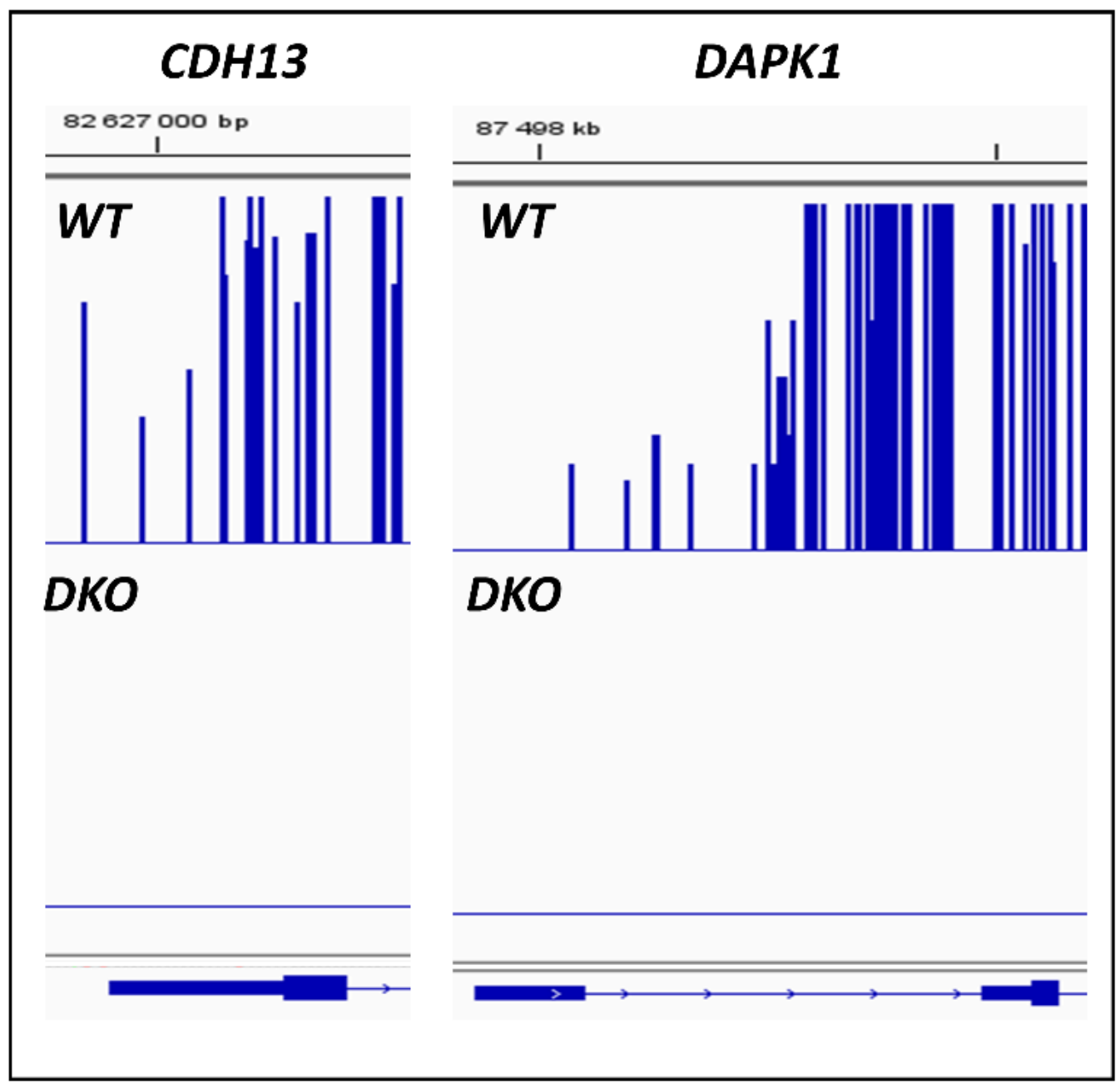

**Figure 4.** Visualisation of the methylation status of *CDH13* and *DAPK1* TSG promoters in wild-type (WT) and DNMT-deficient (DKO) HCT116 colorectal cancer cells. Whole-genome bisulfite sequencing (WGBS) raw data from Blattler *et al.*[62]

Several locus-dependent observations help explain the seemingly contradictory roles of DNA methylation in transcription. While gene promoters primarily regulate protein-coding gene expression, which is repressed when CGIs are methylated, intragenic CpGs serve different functions. Intragenic CpGs are predominantly associated with repetitive DNA elements, such as retroviruses, long interspersed nuclear elements (LINEs), Alu sequences, and transposons, all of which are involved in maintaining genome stability.[44] Methylation of these elements silences their activation, preventing genomic instability while ensuring host gene transcription.[67] Gene bodies also contain CpGs within intragenic promoters that regulate the transcription of non-coding RNAs (ncRNAs) with regulatory functions. DNA methylation impacts these regions, as evidenced by tumour-specific hypermethylation of promoters for transfer RNAs, PIWI-interacting RNAs, and microRNAs such as miR-200s, miR-9, and miR-124.[68] In intragenic regions where CpG methylation is highly variable, such as enhancer sequences or exon-intron junctions, DNA methylation plays more regulatory than transcriptional roles, often mediated by additional epigenetic partners.[44] For instance, studies suggest that intragenic CpG methylation influences alternative splicing during transcription. A proposed mechanism involves the recruitment of the chromatin architectural protein CCCTC-binding factor (CTCF), which interacts with RNA polymerase II to control transcription kinetics, a critical parameter in splicing events.[69] CTCF also binds to unmethylated enhancers, distal regulatory regions that modulate transcription by altering chromatin 3D topology and disrupting enhancer-promoter interactions. As an example, this insulator-like behaviour, dependent on aberrant DNA methylation, is considered responsible for oncogenic transcription signatures.[61]

The functional role of DNA methylation has thus moved beyond a simplistic on/off mechanism of gene expression, where promoter methylation equals silencing and demethylation equals activation, to embrace a multifactorial epigenetic regulatory framework. First, gene transcription can be modulated by different levels of methylation densities on a promoter, resulting in a graded range of stable expression states, overcoming the idea a binary mechanism.[70,71] Then, several observations question a linear relationship between DNA methylation and transcription: 30% of gene promoters lack CGIs,[12] only 16% of CpG sites surrounding transcriptional start sites correlate with gene silencing,[72] and CpG sites at promoters remain unmethylated during embryogenesis regardless of gene transcription status.[61] Additionally, transcription can be silenced by Polycomb proteins even in the absence of promoter methylation.[73] It is therefore evident that discriminating the specific impact of DNA methylation within the broader epigenetic landscape is crucial for understanding how to control this mechanism in pathological contexts. This will not be possible without appropriate tools and technological solutions capable of regulating DNA methylation at specific genomic regions.

## 4. Existing strategies to target DNA methylation

Since the discovery of DNA methylation, considerable efforts have been made to develop compounds capable of control DNA methylation to modulate gene expression. The most explored derivatives have been mainly agents capable of reducing CpG methylation across the genome. Such compounds are known as hypomethylating agents and are non-specific. They were primarily used as anticancer agents to enable the re-expression of TSGs silenced by promoter hypermethylation.[19] However, research has expanded to include many other diseases, such as cardiovascular, infectious, neurodegenerative, and aging-related disorders.[74] Especially, there is growing interest in genetic diseases, including sickle cell disease, for which a first non-nucleoside DNMTi has entered clinical trials (NCT number:

NCT05660265). Below, we depict the classes of compounds with modulatory activity on DNA methylation.

### a) Inhibition of DNA methylation

The most explored strategy to control DNA methylation involves the use of DNMTi. These can be classified into two major categories: nucleoside analogues, which mimic cytosine and are incorporated into DNA after intracellular phosphorylation, and non-nucleoside analogues, which interact with the DNMT catalytic or allosteric pocket without DNA incorporation. In some cases, subtype-selective inhibitors have been developed. The emergence of new concepts in drug discovery has expanded the approaches available to inhibit DNMTs, for example, through multitarget inhibitors and PROTACs.

*Nucleoside analogues*

The first hypomethylating agents described were nucleoside analogues, such as 5-azacytidine (5aza, Vidaza®) and 5-aza-2'-deoxycytidine (5aza-dC, Decitabine, Dacogen™) (**Figure 5**). These compounds act as DNMT substrates and suicide inhibitors. After phosphorylation within the cell, they are incorporated into DNA during replication at random loci. The presence of a nitrogen or substituted carbon at position 5 in nucleoside DNMTi prevents β-elimination of the catalytic cysteine during DNA methylation and the release of the enzyme at the end of the catalytic cycle, leading to the irreversible binding of DNMT to DNA and its proteasomal degradation.[75] This results in progressive passive demethylation over several DNA replications.[76] 5aza and 5aza-dC were initially developed as cytotoxic agents for leukaemia treatment.[77] Their hypomethylating activity via DNMT inhibition at lower doses was later established.[78] In fact, at low, non-intrinsically toxic doses, these compounds induce global and sustained DNA hypomethylation and reactivation of TSGs, such as the cyclin-dependent kinase inhibitor *CDKN2B* and death-associated protein kinase *DAPK*, leading to cell cycle arrest and apoptosis in leukemic cell lines.[79] However, their efficacy is limited due to their short half-life under physiological conditions and rapid degradation by cytidine deaminase.[80,81] 5-fluoro-2'-deoxycytidine (5-F-dC)[82] and 5-aza-4'-thio-2'-deoxycytidine (5aza-T-dC) represent a second generation of nucleoside DNMTi with improved metabolic stability due to their enhanced resistance to cytidine deaminases.[83] Prodrugs of nucleoside analogues have also been developed, such as SGI-110 (Guadecitabine), a dinucleotide composed of decitabine and deoxyguanosine. While deoxyguanosine increases compound stability against deaminases, decitabine is released as the active agent after intracellular dephosphorylation.[84]

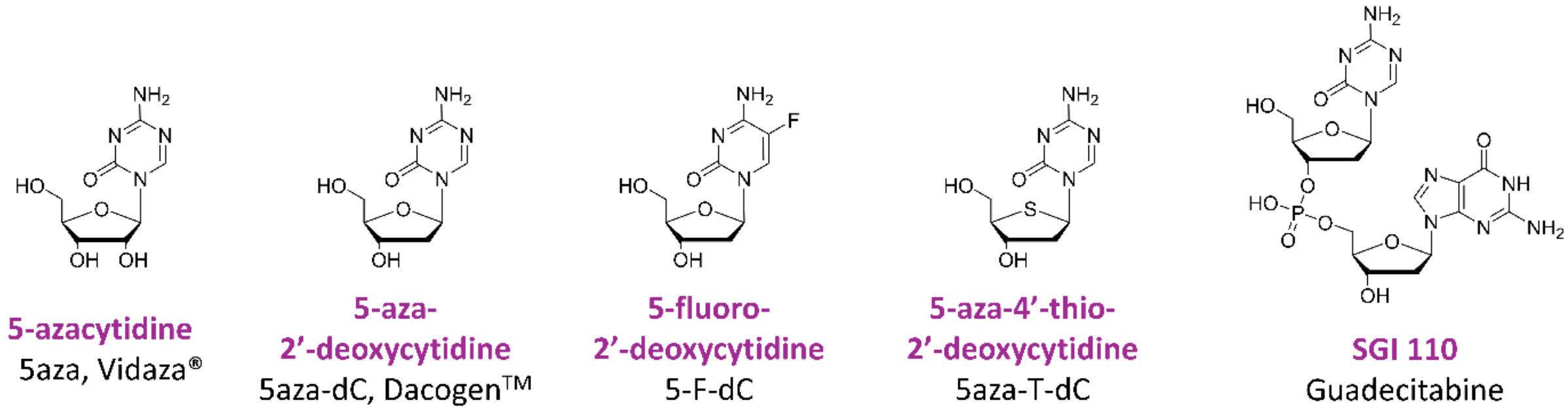

**Figure 5.** Chemical structure of nucleoside DNMT inhibitors.

*Non-nucleoside analogues*

Non-nucleoside DNMTi were discovered as hypomethylating agents that do not incorporate into DNA, thus overcoming the limitations of nucleoside analogues. Several natural products, including curcumin, genistein, and (-)-epigallocatechin-3-gallate (EGCG), have been described as non-nucleoside hypomethylating agents, although their direct interaction with DNMT remains unclear (**Figure 6**). Their effects are often indirect, involving the modulation of DNMT expression or activity through multiple signalling pathways.[74,85] Drug repurposing has led to the identification of hydralazine and procainamide, two cardiovascular agents used clinically, as DNMT modulators, although only weak binding to DNMT has been demonstrated.[86]

Among synthetic compounds, DNA and SAM substrate-competitive inhibitors have been extensively developed (**Figure 6**). RG108 was one of the first non-nucleoside DNMTi identified by in silico screening of the DNMT1 catalytic pocket.[87] RG108 reversibly interacts with the DNMT catalytic pocket through its phthaloyltryptophan scaffold, where the phtalimide moiety binds to the cytidine pocket. Although it induced *p16* and *TIMP* demethylation and transcription reactivation in HCT-116 colorectal cancer cell line, its inhibitory potency toward human DNMT remained modest, limiting its further development.[88] Additionally, different IC50 values were reported for these compounds spanning 1 µM to 390 µM.[88,89] SGI-1027, a quinoline derivative, was also described as a DNA-competitive DNMTi.[90] The original paper describing the molecules has been retracted due to image duplication, but following studies demonstrated its potential as DNMTi, while its analogues showed to inhibit DNMT by DNA intercalation.[91–93] Based on this scaffold, bisubstrate derivatives were designed through rational design to mimic the transition state of the methyl group transfer between SAM (mimicked by the quinazolinone moiety) and DNA cytidine (mimicked by the quinoline moiety).[94] This represents a potent class of compounds capable of decreasing gene promoter methylation in a luciferase reporter cell model[94] and reducing artemisinin-resistant *P. falciparum* infection in mice.[95] However, their development is limited by low water solubility and pharmacokinetic properties.[94] Finally, a high-throughput screening (HTS) campaign followed by hit optimization led to the identification of flavanone scaffolds as potent DNMT3 inhibitors.[96–98] Among them, 3,3-dihalo-2-methoxyflavanones showed high water stability and inhibitory activity. Notably, compound 34b (anti-3-bromo-3-chloro-2-methoxyflavanone) represents the most potent DNMT3 inhibitor developed to date, with submicromolar and anticancer activity.[99]

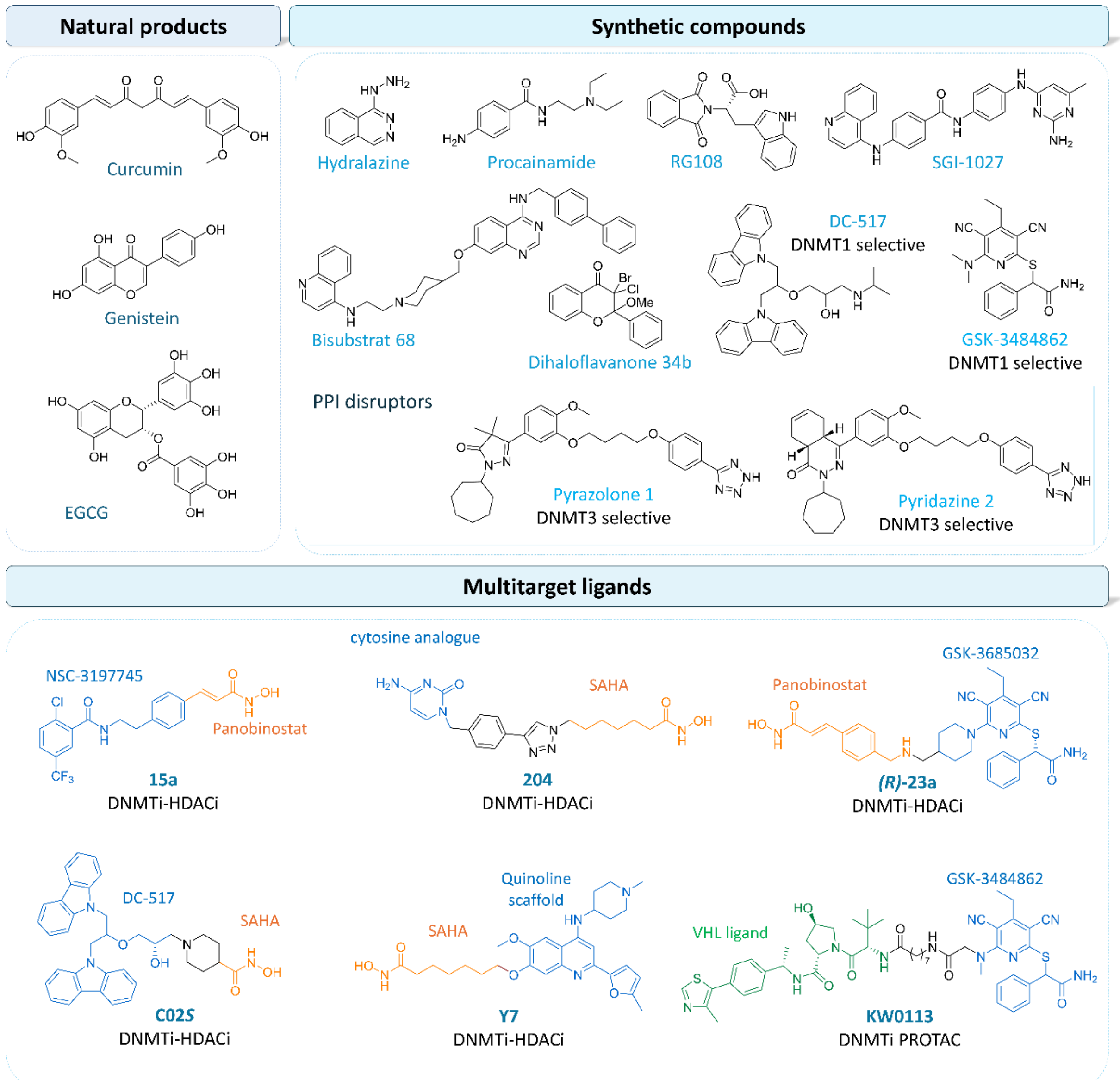

**Figure 6.** Chemical structure of DNMT inhibitors discovered from natural sources or synthetically synthesised, including multitarget probes.

*Subtype- and function-selective DNMT inhibitors*

Efforts to improve subtype specificity have led to the discovery of additional DNMTi. The dicarbazole derivative DC-517 was the first DNMT1-selective inhibitor identified by virtual screening. It binds to both the SAM and cytidine pockets at micromolar IC50 values.[100,101] More recently, a HTS campaign identified dicyanopyridines with promising DNMT1-selective inhibitory activities.[102] In particular, GSK-3484862 and GSK-3685032 stand out for their nanomolar IC50s and solid cellular results, showing antitumor efficacy in 52 hematological cancer cell lines. According to the co-crystal structure analyses, their selectivity toward DNMT1 arises from the intercalation of the dicyanopyridine compound into hemimethylated, but not unmethylated, double-stranded DNA, and its interaction with the catalytic pocket of DNMT1, preventing base flipping and thus blocking cytosine methylation.[103] These derivatives are the first compounds exhibiting *in vivo* DNMT inhibitory activity comparable to that of nucleoside analogues. Due to their lower toxicity, they can be administered at higher doses, inducing a stronger antitumor response than 5aza-dC in MV4-11 and SKM-1 AML xenograft mouse models.[102]

As previously described, DNMT3A/B and DNMT1 interact with distinct partner proteins responsible for differential epigenetic regulation in cells. By exploiting these specific interactions, protein-protein interaction (PPI) disruptors have also been developed to selectively target not only DNMT subtypes but also the functional epigenetic effects associated with specific interactions. For example, pyrazolone and pyridazine derivatives were identified as allosteric DNMT3A inhibitors through the screening of a small chemical library,[104] and their mechanism of action as PPI disruptors was further characterised.[105] These compounds disrupt DNMT3A homo- and heterotetrameric formation, likely acting at the interface of DNMT3A and DNMT3L interaction, reducing overall MTase activity.[104] More recently, a peptide ($EXD^{DNMT1/ELK1}$, structure not disclosed) was developed to disrupt the interaction between DNMT1 and the transcription factor ELK1 using protein-protein interaction prediction by AlphaFold2. $EXD^{DNMT1/ELK1}$ enabled selective demethylation of the diacylglycerol kinase iota (*DGKI*) promoter without affecting the O6-methylguanine methyltransferase (*MGMT*) promoter. In glioblastoma, patients with methylated *MGMT* and unmethylated *DGKI* have a better prognosis, making targeted demethylation of DGKI particularly beneficial for improving therapeutic outcomes.[106] Importantly, by acting only at preferential loci where DNMT1 methylates DNA in association with ELK1, $EXD^{DNMT1/ELK1}$ does not induce genome-wide demethylation, which is promising for glioblastoma therapy with reduced side effects.

DNA methylation can also be modulated by indirectly downregulating DNMT cellular levels using mRNA targeting agents. The antisense phosphorothioate oligodeoxynucleotide MG98 selectively targets the 3'-untranslated region of DNMT1 mRNA through base pairing and blocks its translation, resulting in dose-dependent depletion of DNMT1 without affecting DNMT3, followed by TSG *CDKN2A* re-expression in the HCT116 cancer cell line.[107]

*Novel DNMTi-based probes (Multitarget inhibitors and degraders)*

A promising strategy to treat multifactorial diseases involves the use of multitarget agents. In the context of DNA methylation, this approach offers a valuable alternative to increase the potency of existing DNMTi. Combining pharmacophores in a single molecule does not necessarily increase inhibitory potency against the target enzyme but can enhance antitumor efficacy in cells and *in vivo* compared to drug combinations.[108] The combination of DNMTi and histone deacetylase inhibitors (HDACi) has already demonstrated synergistic effects in vitro and in several clinical trials for haematological and solid malignancies.[109–111] Several DNMT-HDAC hybrid inhibitors have been described (**Figure 6**), linking hydroxamic acid derivatives as HDAC inhibitors to different DNMT inhibitor scaffolds, such as NSC-3197745,[112] nucleoside-based inhibitors,[113,114] quinoline derivatives,[115–117] bicarbazole derivatives,[118] and dicyanopyridine derivatives.[119] Among them, the bicarbazole derivative C02S showed comparable tumour growth inhibition to the combination of SAHA, a clinically approved HDACi, and decitabine in the 4T1 breast tumour xenograft mouse model.[118] The selective DNMT1/HDAC dicyanopyridine compound (*R*)-23a also exhibited similar tumour growth inhibition compared to the combination of GSK-3685032 and SAHA in the MV4-11 AML xenograft model, along with reduced toxicity to non-tumoral hematopoietic cells.[119] Only the quinoline derivative Y7 showed better potency than the combination of SAHA and decitabine in the MMTV-PyMT transgenic mouse model of spontaneous breast cancer, associated with greater tumour volume reduction and notable suppression of metastasis.[117]

Proteolysis-targeting chimeras (PROTACs) are bivalent probes that simultaneously target a protein of interest and an E3 ubiquitin ligase. This is an efficient and recognised strategy to hijack the protein of interest for proteasomal degradation through E3 ligase proximity-induced ubiquitination. Compound KW0113 (**Figure 6**), based on the DNMT1-selective inhibitor GSK-3484862 and a VHL E3 ligase ligand, is the only PROTAC targeting DNMT1 degradation described so far. KW0113 exhibits selective, time- and submicromolar dose-dependent degradation of DNMT1 in MV4–11 and MOLM-13 AML cell lines.[120] However, recent studies have revealed that GSK-3484862 itself acts as a DNMT1 degrader.[121,122] Another DNMT1 molecular glue has been serendipitously discovered and demonstrated higher efficacy in UHRF1-mediated DNMT1 degradation in AML cells.[123] DNMT1 degraders induce global hypomethylation similarly to traditional and next-generation DNMTi. While their anticancer efficacy is well described, no information on any differential methylation patterns induced by degraders *vs* DNMTi is available to date. Further investigation could provide interesting insights for novel therapeutic strategies.

*DNMT inhibitors in the clinic*

To date, only a few nucleoside DNMT inhibitors have reached clinical approval. 5aza (**Figure 5**) was first approved by the FDA for the treatment of myelodysplastic syndromes (MDS) in 2004 and later extended to acute myeloid leukemia (AML) in 2020 and juvenile myelomonocytic leukemia (JMML) in 2022. 5aza-dC (**Figure 5**) was approved for the treatment of MDS in 2006 and AML in 2018.[81,111] These compounds are also involved in numerous ongoing clinical trials, either alone or in combination with chemotherapies, radiotherapies, or other epigenetic drugs. However, their clinical benefit is limited to the treatment of haematological malignancies and palliative administration at low doses due to their high off-target toxicity, lack of selectivity, and chemical and metabolic instability. One strategy to improve the pharmacokinetic properties of nucleoside analogues is to co-administer them with cytidine deaminase inhibitors, as exemplified by the oral combination of 5aza-dC and cedazuridine, which was approved in 2020 for the treatment of MDS.[124]

Gastrointestinal toxicity and myelosuppression are the most common adverse effects reported 5aza and 5aza-dC.[81] Despite an average survival of only 13 months, more than 30% of AML/MDS patients treated with azacytidine experienced myelosuppression, and more than 80% of patients treated with decitabine developed neutropenic fever and more than 10% developed infections.[125] In mouse models of solid tumours, including breast, colon, and lung epithelial cancers, low doses of 5aza and 5aza-dC regulates anti-tumour gene expression without causing cell death.[79] At higher doses, drug systemic toxicity was not sustainable.[126,127] While this first discouraged their clinical application, preclinical studies suggested that 5aza and 5aza-dC could reverse DNMT-dependent chemoresistance to cetuximab and platinum-derived agents.[128]

The second-generation nucleoside 5-F-dC (**Figure 5**) is currently in phase II clinical trials in combination with tetrahydrouridine, a cytidine deaminase inhibitor, for advanced non-small cell lung cancer, breast cancer, bladder cancer, or head and neck cancer (NCT00978250),[129] and in phase I for refractory solid tumours (NCT01534598).[130] 4'-thio-dC (**Figure 5**) has also recently entered several phase-I clinical trials for the treatment of solid tumors (NCT02423057), MDS (NCT03366116), and AML (NCT04167917).[83,131] The prodrug guadecitabine (**Figure 5**) has been evaluated in multiple phase II and III clinical trials for the treatment of AML (NCT03603964) and MDS (NCT02131597), gastrointestinal stromal tumours (NCT03165721), and Philadelphia-negative myeloproliferative neoplasms, a hematopoietic stem cell disease (NCT03075826). Although it demonstrated good pharmacokinetics and sustained DNA hypomethylation, its benefits compared to 5aza-dC treatment have yet to be demonstrated.

The antisense oligonucleotide MG98 entered phase I clinical trials to evaluate its pharmacokinetic and pharmacodynamic properties[132] and its efficacy in advanced solid tumours,[133] which was extended to phase II for renal carcinoma.[134] Although the drug was well tolerated, no antitumor effect was observed in phase II, likely due to the accumulation of the antisense oligonucleotide in the liver.[134] Finally, in 2023, the DNMT1-selective inhibitor GSK-4172239D (undisclosed structure) became the first non-nucleoside DNMT inhibitor to enter clinical trials for the treatment of sickle cell disease (Phase I, NCT05660265). In this context, DNMT inhibition promotes the re-expression of foetal haemoglobin, which replaces the defective adult haemoglobin responsible for the disease, without inducing significant toxicity.[135] However, this study was terminated, likely for strategic business considerations.

Overall, despite various efforts to develop new DNMTi, the clinical landscape remains largely dominated by first-generation nucleoside analogues. The development of more selective, stable, and less toxic compounds, particularly non-nucleoside derivatives, still represents a major challenge. Emerging strategies to target selective DNMT subtypes or pathway-specific DNA methylation remain at early-stage development, and approaches to control DNA methylation at desired genomic loci using small molecules are still underexplored.

**b) Inhibition of DNA demethylation**

While DNMT inhibition has been extensively studied over the past four decades, the more recent discovery of the TET enzymes has opened new avenues for epigenetic modulation. Targeting TET proteins represents a promising therapeutic strategy since they have emerged as key regulators in a wide range of diseases, including cancer, inflammatory diseases, but also cardiovascular,[136] and neurological[137] disorders. In cancer, TET enzymes are generally considered to have tumour-suppressive functions, in contrast to DNMTs. However, their role is more tumour dependent, and their deregulation can contribute to DNA methylation imbalance and tumour progression.[138,139] For example, mutations in the gene coding for TET2, the predominant TET isoform in haematopoietic cells, are among the most frequent in myeloid neoplasms and result in a loss-of-function. As a compensatory mechanism, TET1 and TET3 maintain a basal level of 5hmC, essential for cell survival. TET2 mutated haematopoietic cells are therefore more sensitive to TET inhibitors, causing lethality.[140] The role of TET enzymes has also been explored in immune and inflammatory disorders. In fact, TET1/2 inhibition modulates the immunomodulatory function of mesenchymal stem cells (MSCs) by inducing promoter hypermethylation and consequent silencing of DKK-1, a component of the Wnt/β-catenin signaling pathway, and reducing inflammation through T cell apoptosis.[141] However, to date, the identification of TET-specific ligands remains relatively underexplored, despite the emergence of several promising scaffolds. TET enzymes can be targeted through multiple strategies, including the development of nucleoside analogues and co-factor mimicking agents, like 2-oxoglutarate (2-OG) and Fe(II)-chelating derivatives.[138]

The first nucleoside analogues targeting TET enzymes were designed based on the crystal structure of the TET2-DNA complex (PDB: 9HXV). Exploration of methyl bioisosteres led to the identification of a series of N-biphenyl-5-chlorocytosine derivatives, among which Bobcat339 (**Figure 7**) exhibited weak inhibitory activity against TET1 and TET2, with an IC50 no lower than 30 µM and no activity against DNMT3A. Despite its modest potency, Bobcat339 significantly reduced 5hmC levels in HT-22 hippocampal neurons at 10 µM, positioning this scaffold as a valuable starting point for the development of nucleoside-based TET inhibitors.[142]

IOX1, a quinoline derivative, acts as a broad-spectrum 2-OG-competitive inhibitor, inhibiting TET2 at micromolar concentrations.[143] This compound represents a potent scaffold for TET inhibition, particularly through the development of sulfonate-substituted quinoline derivatives. For instance, compound 2a, identified via HTS, exhibited an IC50 in the micromolar range against TET2, owing to its ability to chelate Fe(II) in the catalytic site.[144] A subsequent study further optimised the IOX1 scaffold, yielding a carboxylic acid analogue (compound 2b) that inhibits TET2 with an IC50 of 0.5 µM. However, its selectivity across TET-sensitive enzymes remains to be evaluated.[143]

In cells, 2-hydroxyglutarate (2-HG), an endogenous metabolite, acts as a competitive inhibitor of 2-OG and serves as another scaffold for TET inhibition.[145] Notably, TETi76, a 4-methylene derivative of 2-HG, inhibits TET enzymes at micromolar concentrations, with a preference for TET1. In addition to its enzymatic inhibition, TETi76 demonstrated potent clonal suppression of TET2-mutated leukemic cells in a xenograft model, while sparing normal cells, thereby highlighting the therapeutic potential of TET inhibitors.[146]

The development of Fe(II) chelators as TET inhibitors presents a unique challenge, as these compounds may exert indirect effects by sequestering intracellular iron. For example, C35, an anthocyanidin derivative, was identified through structure-based screening targeting TET2 and inhibits all three TET enzymes at micromolar concentrations. While C35 effectively reduced 5hmC levels, it did not significantly alter 5mC levels or modulate the expression of genes involved in the BMP–SMAD–ID pathway, which plays a critical role in cellular differentiation.[147] This compound has been used as a molecular tool to investigate the functional roles of TET enzymes in cell differentiation and reprogramming.

Overall, the development of TET inhibitors remains in its early stages but is rapidly emerging as a valuable strategy to modulate active DNA demethylation and elucidate the functional consequences of 5mC oxidation in epigenetic regulation. Future efforts should focus on improving potency, selectivity, and cellular permeability, while minimising off-target effects to fully harness the therapeutic potential of TET modulation.

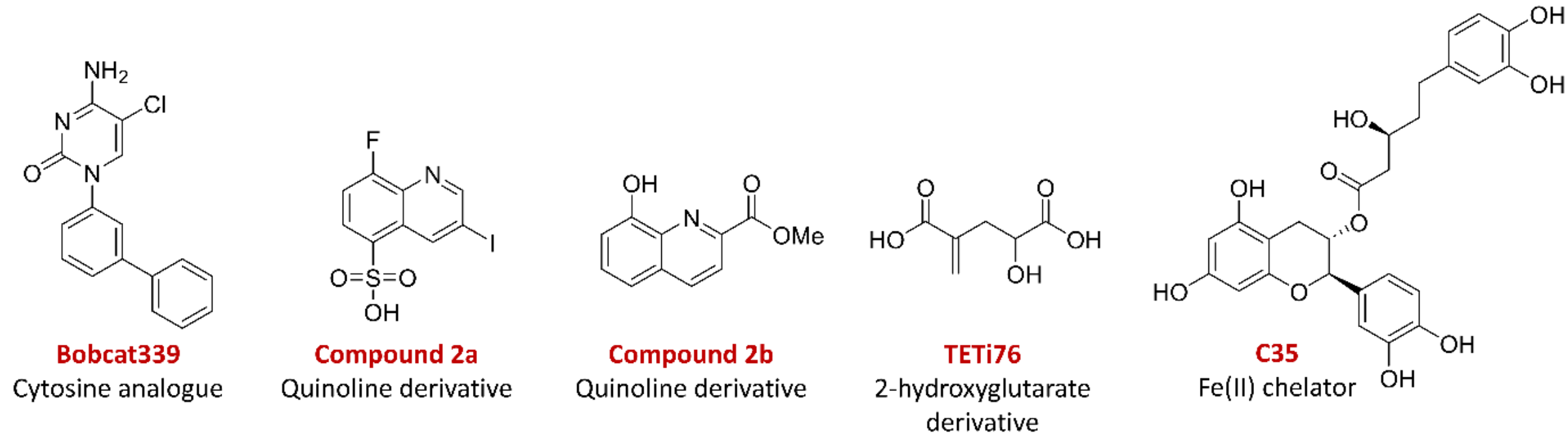

**Figure 7.** Chemical structure of TET inhibitors.

### c) Direct chemical modification of DNA methylation

The strategies previously described to modulate DNA methylation primarily target DNMTs within cells. However, certain chemical compounds can directly methylate DNA through non-enzymatic alkylation, as exemplified by TMZ (**Figure 8**), discussed earlier. Direct chemical DNA methylation is rarely nucleobase-specific, although certain preferences may emerge depending on the reactivity of nucleobases toward the methylating agent. For instance, TMZ exhibits a preference for methylating guanine at positions N7 and O6, and adenine at position N3.[5] Dacarbazine (DTIC, **Figure 8**) represents another methylating agent with similar activity,[148] while novel TMZ derivatives have been developed to modulate its pharmacokinetic properties, though their DNA methylation mechanism remains unchanged.[149]

Beyond triazenes, methyl methanesulfonates have been explored as alternative methylating agents, primarily in anticancer research. While methyl methanesulfonate itself preferentially methylates guanosine at the N7 position,[150] derivatives incorporating DNA-binding motifs (**Figure 8**) that interact with adenine- and thymine-rich regions have shown increased adenine N3 methylation.[151] Methylation at this position is considered a highly cytotoxic DNA lesion with low mutagenicity. Overall, adenine and guanine are the most relevant methyl acceptors for these chemical donors, explaining their use in inducing DNA adducts and strand breaks, particularly in combination with inhibitors of DNA repair systems (e.g., PARP inhibitors) to trigger cancer cell apoptosis.[151] Since cytosine is not a primary target, these agents do not exert epigenetic activity through DNA methylation. However, developing methods to induce cytosine methylation chemically and exert epigenetic effects in cells remains an intriguing and unexplored avenue, especially given the current lack of small-molecule DNMT activators.

TMZ DTIC Methyl methanesulphonate DNA binding motif

**Figure 8.** Reported chemical derivatives directly methylating DNA in cells.

Nevertheless, recent advances have identified molecules capable of chemically oxidising specific nucleobase methyl groups, such as N3-methyladenosine and 5mC, to label and identify these epigenetic modifications in oligonucleotide pools using photochemical radical reactions.[152,153] Based on these early studies, the Daumann and Carell groups designed an Fe(IV)-based compound that selectively oxidised 5mC in oligonucleotides with minimal hydrolytic side products, mimicking TET-mediated oxidation.[154] This mild oxidation converted 20% of 5mC nucleosides into 5-carboxy-2'-deoxycytidine. When 5mC was incorporated into an oligonucleotide sequence, the conversion rate dropped to 7%, with the primary products being 5hm) and 5fC. More recently, the Balasubramaniam group developed a photochemical reaction using sodium decatungstate ($Na_4W_{10}O_{32}$) as a photocatalyst and N-fluorobenzenesulfonimide (NFSI) as a single-electron oxidant.[155] This reaction was highly selective for 5mC nucleotides over thymidine and did not produce unwanted by-products such as methyluridine. Translating such chemical strategies in cells could herald a new era for direct epigenetic editing without perturbing enzymatic pathways. However, given the variety of oligonucleotides in the cytoplasm and nucleus, this remains a significant challenge.

## 5. Gene-selective strategies to target DNA methylation

The DNA methylation modulators presented above are not selective for specific genes, as they lack elements capable of targeting determined genomic loci. Any selectivity observed with DNMTi is likely due to the differential accessibility of DNMTs at specific genomic conditions, including nucleosome positioning and heterochromatin formation. These effects would be even more pronounced when comparing non-nucleoside DNMTi with nucleoside analogues, which incorporate into DNA independently of chromatin compaction. On the other hand, DNMT PPI disruptors are not gene-selective but exhibit functional selectivity: by disrupting the interaction of DNMT with a protein partner, they act only on the methylation marks where that partner is involved, without inducing global genomic demethylation.[106] The selectivity provided by PPI disruptors depends on the physiological role of the DNMT-protein interaction and cannot be modulated for any targeted gene. To achieve gene-specific targeting, a DNA-binding motif is necessary.

### a) DNA-binding motifs for gene-selective targeting

N-methylpyrrole and N-methylimidazole polyamides (PIPs) were the first chemical entities developed to recognise specific regions in the minor groove of DNA (**Figure 9**).[156] By altering the sequence of these heterocycles, scientists could design PIPs targeting different genes relevant in cancer. Moreover, by coupling these PIPs with HDACi, these tools could modulate chromatin compaction on determined gene sequences.[157] To our knowledge, no examples in the literature describe the use of DNMTi with PIPs, likely due to the lower activity and commercial availability of DNMTi compared to HDACi before the discovery of GSK ligands.

The complexity of designing chemical DNA-binding motifs like PIPs confined the field of DNA methylation editing to molecular biologists by leveraging gene-specific protein-DNA binders such as $Cys_2His_2$ zinc finger domains (ZFDs) and transcription activator-like effectors (TALEs). ZFDs are small protein domains, typically less than 170 amino acids in length (**Figure 10A**). Approximately 50% of human transcription factors contain this domain to bind their targeted promoters. In the field of DNA methylation editing, ZFDs have been genetically fused with DNMT3A and expressed in MCF7 breast cancer cells to induce methylation of the *SOX2* oncogene. This resulted in a successful reduction of *SOX2* mRNA and protein levels by 60% and 80%, respectively.[158] ZFDs are low in immunogenicity for therapeutic applications, but their engineering is challenging due to the lack of a proper design code to target every desired gene. However, progress in artificial intelligence is providing promising perspectives in this direction.[159]

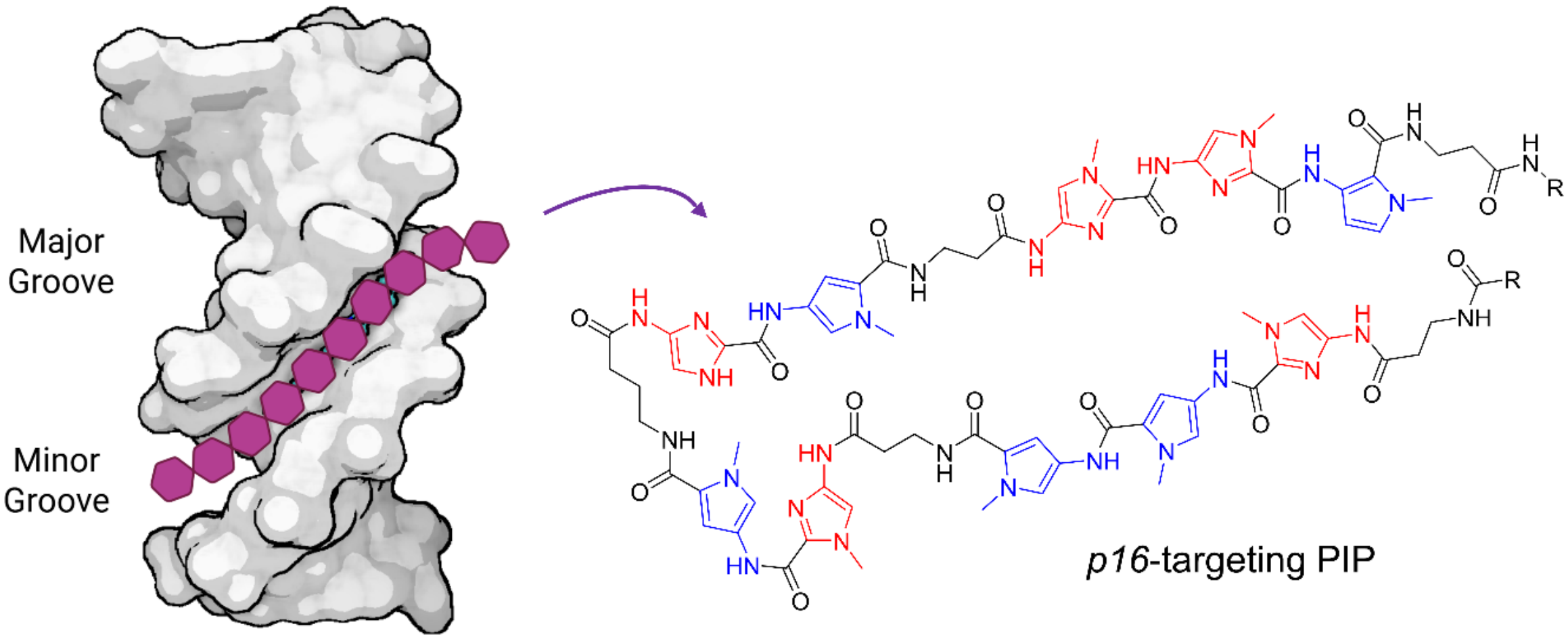

**Figure 9.** Structure of a N-methylpyrrole (blue) and N-methylimidazole (red) polyamides (PIP) targeting *p16* tumour suppressor gene and its interaction in the minor groove of DNA (purple).

TALEs are small bacterial proteins composed of 12 to 27 repeats of 34 amino acids that are polymorphic at positions 12 and 13.[160] These positions are defined as "repeat-variable-di-residues" and determine DNA binding specificity (**Figure 10B**). Each amino acid pair, in fact, preferentially recognises one of the four nucleotides. Thus, TALEs are theoretically easier to engineer compared to ZFDs. Although TALEs have been used to activate gene expression using activator and repressor domains, such as VP64 and SID, respectively, there are only two examples of TALEs designed specifically for DNA methylation control. TALEs were fused to catalytic DNMT3A-3L to target different loci of the TSG *p16* in primary human fibroblasts, reducing their proliferative phenotype.[161] Another study described the use of a similar strategy to target the mouse *Dlk1-Meg3* locus in βTC6 β-cells involved in type 2 diabetes mellitus insulin resistance.[162] However, the design of this construct was poorly characterised, making it difficult to draw any conclusions about the efficacy of TALE-DNMT in modulating DNA methylation in a gene-selective manner.

The advent of CRISPR/Cas9 technology for gene editing has completely transformed the landscape. Cas9 is a bacterial DNA-binding protein that recognises a targeted genomic sequence by loading a complementary single-guide RNA (sgRNA). The recognition mechanism is based on Watson and Crick base pairing, ensuring great reliability and versatility. While Cas9 possesses nuclease activity that cleaves the targeted DNA locus, nuclease-deactivated Cas9 (dCas9) is employed in epigenetic editing.[163] dCas9 contains two mutations, D10A and H840A, in the HNH and RuvC-like nuclease domains, respectively, preventing double-strand DNA cleavage.[164] Similar to ZFDs and TALEs, dCas9 has been fused to several transcriptional activators (e.g., p65, VP64) and repressors (e.g., KRAB, MeCP2).[165–167] These promising tools represented a significant breakthrough in the field and were further evolved to enhance their impact on gene expression. The most recognised development is the SunTag technology, a polypeptide containing up to 24 repeats of an epitope recognised by single-chain variable fragment (scFv) antibodies (**Figure 11A**).[168] SunTag-dCas9 co-transfected with scFv-VP64 recruited multiple copies of VP64 and enhanced the mRNA levels of the cell-cycle inhibitor *CDKN1B* by 330% (with the best sgRNA used) in K562 leukaemia cells, significantly reducing cell growth. For comparison, the dCas9-VP64 fusion protein increased *CDKN1B* mRNA levels by only 28%, with little impact on cell growth. Finally, novel combinatorial strategies coupling dCas9 with small molecules have recently been described. The Di Antonio group designed a dCas9 tool conjugated with a DNA G-quadruplex binder through the HaloTag technology to selectively control the expression of the *MYC* oncogene.[169] dCas9 was also conjugated with bromodomain inhibitor JQ1, enabling gene-selective recruitment of BDR4 and consequent *IL1RN* and *NTF3* gene re-expression,[170] and the dCas9-FKBP fusion protein was paired with bivalent compounds comprising an FKBP ligand and recruiters of endogenous transcriptional activators, such as BRD4, BRPF1, and CBP.[171]

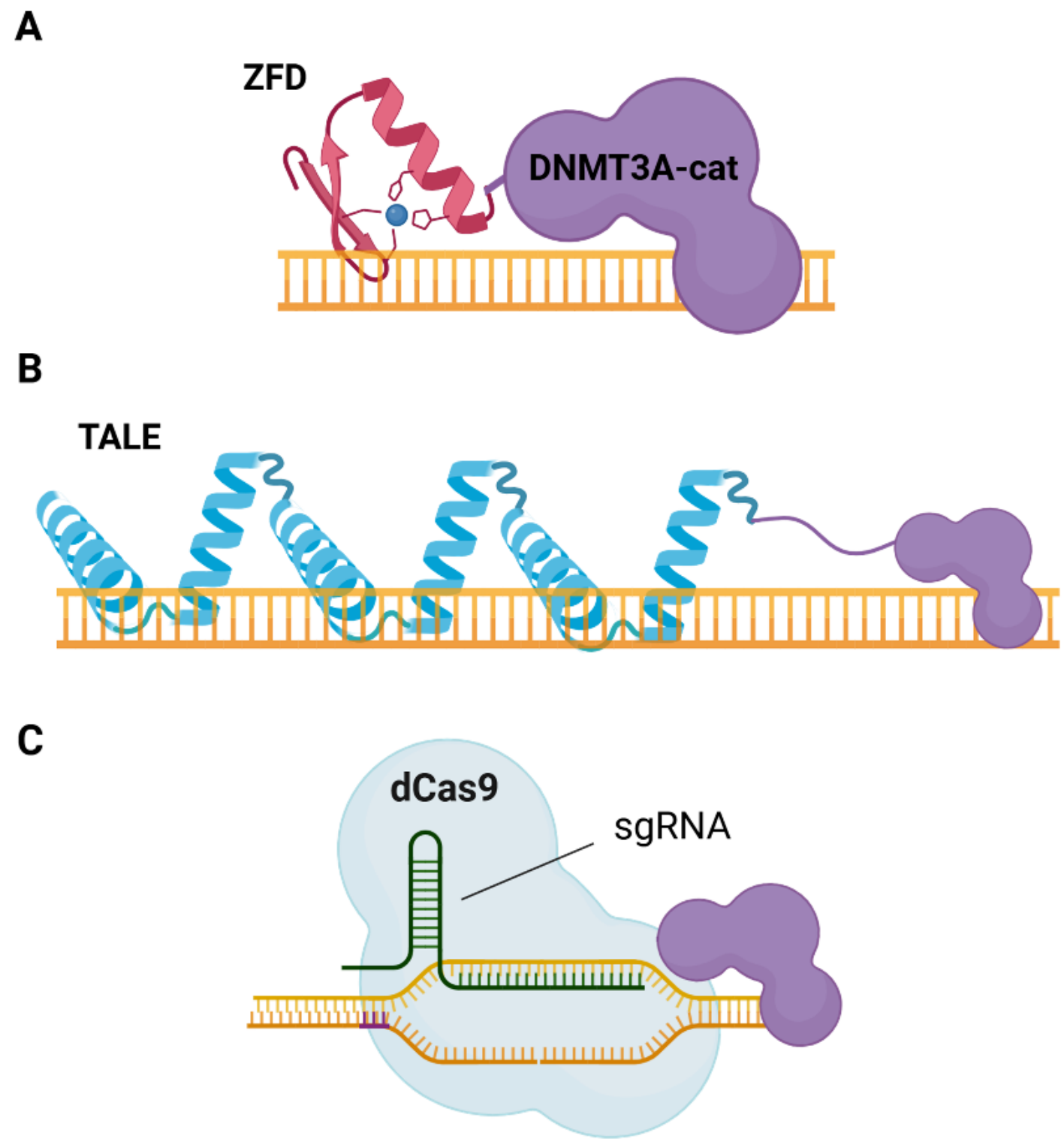

**Figure 10.** Representation of gene-selective tools developed to control DNA methylation: zinc finger domain (ZFD), TALE, and dCas9 fused to the catalytic domain of DNMT3A (DNMT3A-cat).

#### b) dCas9-mediated gene-selective methylation

The dCas9 strategy was also applied to modulate DNA methylation, initially by fusing DNMT3A as an effector protein (**Figure 10C**).[172] By targeting the tumour suppressor gene *p21*, dCas9-DNMT3A induced methylation of 50% of alleles without impacting global methylation. As with similar Cas9-based applications, the choice of sgRNA was critical for achieving significant methylation modification of the desired locus and subsequent gene repression. This represents the only constraint to the versatility of the dCas9 system, which is easily overcome by the straightforward synthesis of sgRNAs. Several examples of dCas9-DNMT3A have been reported, leading to important conclusions about their efficacy. For instance, fusing dCas9 with only the catalytic domain of DNMT3A (DNMT3A-cat) resulted in more efficient but less selective methylation.[173] Moreover, dCas9-DNMT3A-cat suffered from lower nuclear translocation. An efficient approach to increase methylation efficacy without significantly affecting global methylation was to fuse dCas9-DNMT3A with DNMT3L,[174] which proved even more efficient than employing SunTag-dCas9 with scFv-DNMT3A.[175] The loss in efficacy when using the SunTag system for DNA methylation may be due to the high steric hindrance around the targeted gene caused by the recruitment of multiple copies of DNMT3A. Additionally, an overcrowded environment could hamper partner protein recognition, potentially biasing DNA methylation results.

#### c) dCas9-mediated gene-selective demethylation

dCas9 was also fused to TET1 to oxidise 5mCs, inducing selective gene demethylation of the *BDNF* promoter and enhancing its expression in mouse cortical neurons.[176] The dCas9-TET1 system was further developed into a SunTag-based system using scFv-TET1 (**Figure 11A**), achieving more than 90% demethylation in 4 of 9 tested loci.[177] An alternative to the dCas9-TET1 SunTag system was developed by leveraging the plasticity of sgRNA. In fact, it is possible to insert additional RNA elements into the scaffold RNA structure of sgRNAs (**Figure 11B**).[178] The additional elements tested are MS2 bacteriophage coat protein binders. MS2 was then fused with TET1 and co-transfected into HEK-293, HeLa, and SH-SY5Y cell lines together with dCas9 and modified MS2-binding sgRNAs, leading to the upregulation of targeted genes *RANKL*, *MAGEB2*, and *MMP2*. A similar approach was reported as a proof-of-concept for DNMT1 catalytic inhibition on the *RANKL* gene promoter in HEK-293T cells.[179] In this case, dCas9 was expressed with a sgRNA containing an R2-stemloop with high affinity for DNMT1. The R2-stemloop trapped the enzyme around the targeted gene, diluting its methylation over time. A fused system comprising dCas9, DNMT3A/3L, and the gene repressor ZNF10 KRAB successfully resulted in durable gene silencing after 15 months post-transfection, corresponding to approximately 450 cell cycles in 38 out of 39 clones.[180] Gene silencing was reversed using different combinations of transcriptional activators fused to dCas9-TET1, with dCas9-TET1-p65-Rta giving the highest gene reactivation after 2 days post-transfection, an effect maintained until the 28-day endpoint.

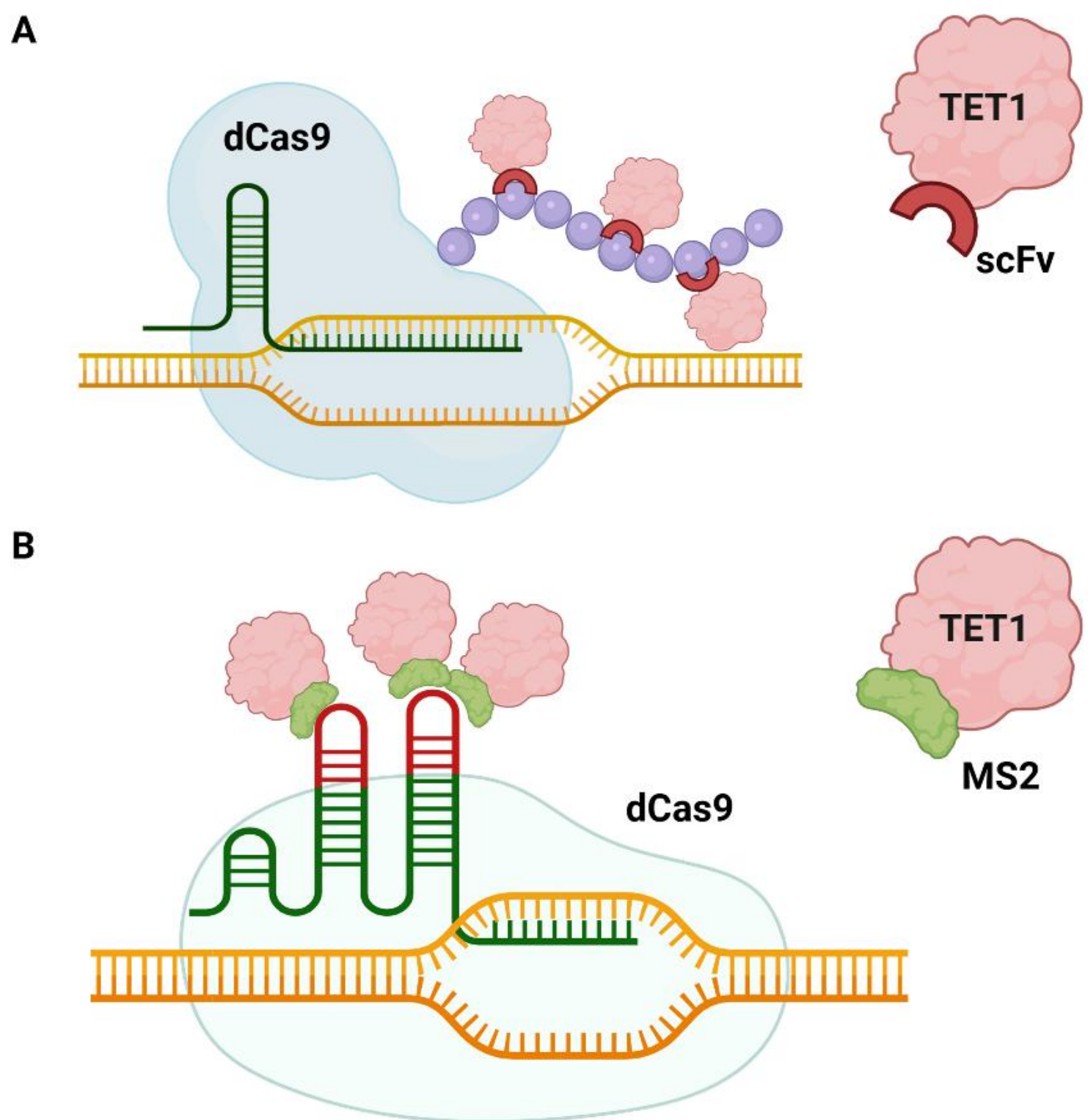

**Figure 11. A)** Representation of the SunTag-dCas9 technology: a polypeptide containing epitopes for scFv is fused to the dCas9, allowing the recruitment of transfected scFv-tagged TET1 to inhibit demethylation at specific loci. **B**) The modification of the sgRNA scaffold with RNA elements recognising the bacteriophage coat protein MS2 allows the gene-selective recruitment of MS2-tagged TET1.

### d) Potential clinical translation of dCas9-based gene-selective DNA modulators

*In vivo* translation of dCas9-based DNA methylation modulators has recently been explored, providing solid foundations for their potential future therapeutic applications. Two cAMP response element (CRE)-inducible dCas9-DNMT3A-cat and dCas9-TET1 transgenic mice were developed and shown to methylate the *Psck9* promoter and reduce LDL cholesterol after appropriate sgRNA-loading adeno-associated virus 9 (AAV9) retro-orbital injection.[181] Furthermore, intracerebroventricular injection of AAV9 carrying *MECP2*-targeting sgRNA reactivated the silenced wild-type *MECP2* allele from the inactive X chromosome in a Rett syndrome mouse model with a loss-of-function mutation of MECP2.

Although the ground-breaking impact of dCas9-based gene-selective fusion systems in modulating DNA methylation, these systems still require extensive studies to assess their feasibility for therapeutics. Their first limitation is the need for genetic manipulation, which raises ethical concerns. Even though the field of protein delivery is advancing rapidly, no protein delivery system has yet been commercialised or clinically approved, meaning that transfection of genetic material remains the most effective method.[182] Additionally, dCas9 is a bacterial protein that triggers extensive innate and adaptive immune responses in mice and potentially in humans.[183–185] Off-target effects are a major theoretical concern with dCas9-based tools. Genome-wide methylation analysis showed that dCas9-DNMT3A-cat targeting the *BACH2* locus in HEK293T and MCF7 cells significantly methylated 32,665 and 10,881 200-bp tiles, respectively.[186] When targeting *C9orf72* in HEK293T in an Illumina Epic Methylation Array, the differentially methylated tiles were 26,125 ($p < 0.01$) or 67,246 ($p < 0.5$).[187] Compared to dCas9-DNMT3A or DNMT3A-cat alone, DNMT3A-based SunTag system showed the most gene-specific results with limited off-target effects compared to even after next-generation sequencing evaluation.[188] Similarly, *in vivo* dCas9-DNMT3A and dCas9-TET1 exhibited only a few off-target effects depending on the number of sgRNAs used to target different loci simultaneously.[181] Galonska et al. suggested a few aspects that may contribute to the off-target effects of dCas9-based systems, such as that dCas9-methyltransferases show ubiquitous nuclear activity in embryonic stem and somatic cells, both in the presence and absence of sgRNAs[186] and that on-target *vs.* off-target accumulation is faster only in a few selected loci. These conclusions align with previous observations of gene methylation using a catalytically deactivated dCas9-DNMT3A in HEK-293 cells, suggesting that transfection of exogenous effector proteins may scaffold with endogenous DNMTs, resulting in uncontrolled methylation of non-targeted genes.[189] In fact, it is well known that DNMT3A and DNMT3B are prone to oligomerisation, which impacts their methylation activity.[190] Nevertheless, the potential impact of off-target effects on the treatment of any given disorder must be carefully evaluated on a case-by-case basis to determine the risk-benefit ratio. Given that the *in vivo* application of gene-selective systems remains in its early stages, comprehensive data on their safety and efficacy are not yet available. As a future perspective, novel tools based on combinatorial strategies, as presented in section 5a, may achieve more clinical translation-friendly opportunities to control DNA methylation in a gene-selective fashion due to their different nature and mechanism of action compared to current systems.

## Conclusions

DNA methylation embodies a dual paradox in biology and medicine: it serves as both a toxic alkylating mechanism harnessed by chemotherapeutic agents to induce cancer cell death and a sophisticated epigenetic regulator essential for cellular physiology. In this review we focused on the pathophysiological role of DNA methylation. Over the past decades, our understanding of DNA methylation has evolved from a simplistic model of gene silencing, where promoter methylation equated to transcriptional repression, to a nuanced, locus-specific mechanism intricately connected with histone modifications, chromatin remodelling, and non-coding RNAs. This complexity underscores the need for tools capable of dissecting its context-dependent roles across diverse pathophysiological scenarios.

While traditional DNMTi and emerging non-nucleoside analogues have expanded our pharmacological arsenal, they still fall short of the precision required to elucidate the genomic locus-specific functions of DNA methylation, especially in an era where next-generation sequencing and omics technologies provide unprecedented resolution for epigenetic analyses. In this review, we have examined the mechanisms and roles of DNA methylation in epigenetic regulation, from its enzymatic modulation by DNMTs and TET proteins to its interplay with chromatin architecture. We have also evaluated the current landscape of DNA methylation modulators, ranging from first-generation nucleoside analogues to innovative CRISPR-dCas9 fusion systems and PPI disruptors. These advanced tools, such as dCas9-DNMT3A, dCas9-TET1, and PPI-based inhibitors, offer promising avenues for achieving locus-specific modulation, thereby minimising off-target effects and enhancing therapeutic precision. Yet, challenges remain, particularly in optimising delivery systems, mitigating immune responses, and addressing off-target activity, issues that must be carefully evaluated to ensure the safety and efficacy of these approaches.

Looking ahead, the development of precise, gene-selective tools represents a transformative opportunity for both cancer research and therapeutic innovation. By integrating computational design, high-throughput screening, and novel delivery strategies, we can refine these tools to target specific genomic loci with minimal collateral damage. Such advancements will not only deepen our understanding of DNA methylation's role in health and disease but also pave the way for personalised epigenetic therapies tailored to individual genomic landscapes. Ultimately, the ability to modulate DNA methylation with locus resolution holds the potential to revolutionise the treatment of cancer, neurodegenerative disorders, and genetic diseases.

## Acknowledgments

Some figures in this manuscript were created entirely or partially on BioRender. The authors acknowledge the support of the Centre National de la Recherche Scientifique (CNRS) and the administrative staff of the Institute of Biomolecules Max Mousseron (IBMM).

## Authors Biographies

**Dr Julie Gilbert** obtained her PhD in Medicinal Chemistry from the University of Montpellier in 2024, where she developed multitarget strategies to modulate epigenetic mechanisms deregulated in cancer. She is currently a postdoctoral researcher in the Department of Chemistry at the University of Oxford, where she investigates the role of reactive metabolites as novel signalling regulators and drivers of disease.

**Dr Francesco Calzaferri** is a CNRS researcher at the Institute of Biomolecules Max Mousseron (IBMM) in Montpellier since January 2026. After obtaining his PhD in Pharmacology from the University Autónoma of Madrid (UAM) in 2021 through the Marie S. Curie ITN programme “Purines DX”, he dedicated his research in epigenetic targeting and drug delivery systems. His research currently focuses on the discovery and optimisation of novel DNA and histone methyltransferase inhibitors, as well as the development of gene-selective chemical tools with therapeutic potential.